\documentclass[JHEP,12pt]{article}
\usepackage{booktabs}

\usepackage{scalerel}

\newcommand{\mzero}{\includegraphics[valign=m,scale=0.15]{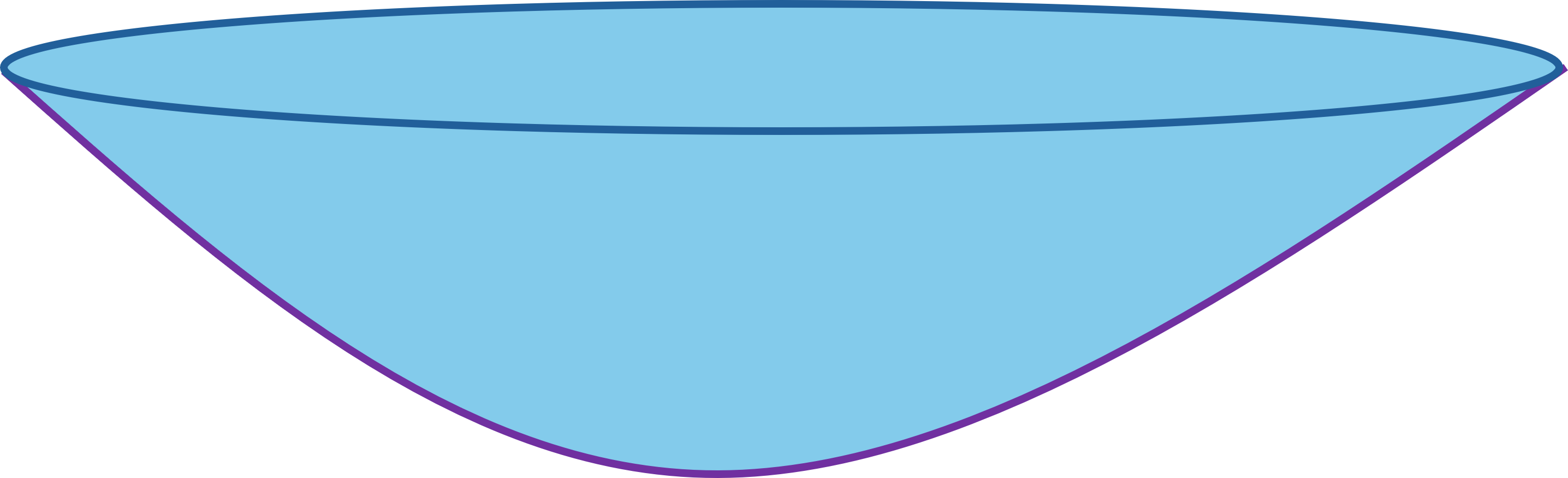}}
\newcommand{\mone}{\includegraphics[valign=m,scale=0.15]{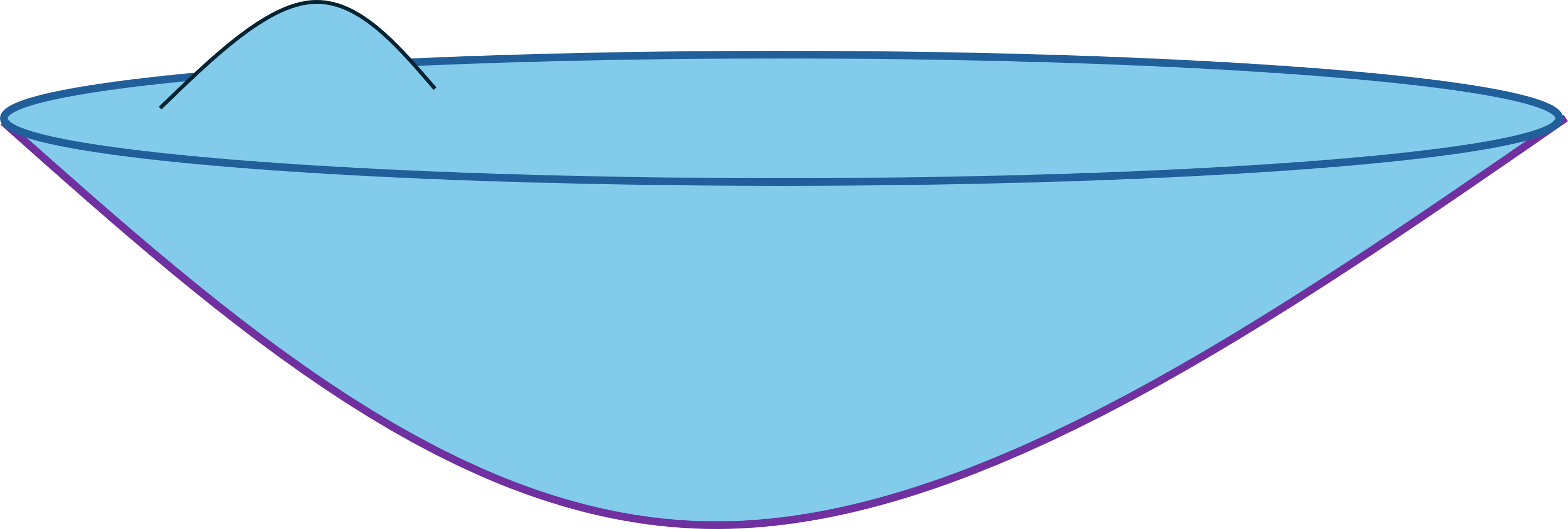}}
\newcommand{\mtwo}{\includegraphics[valign=m,scale=0.15]{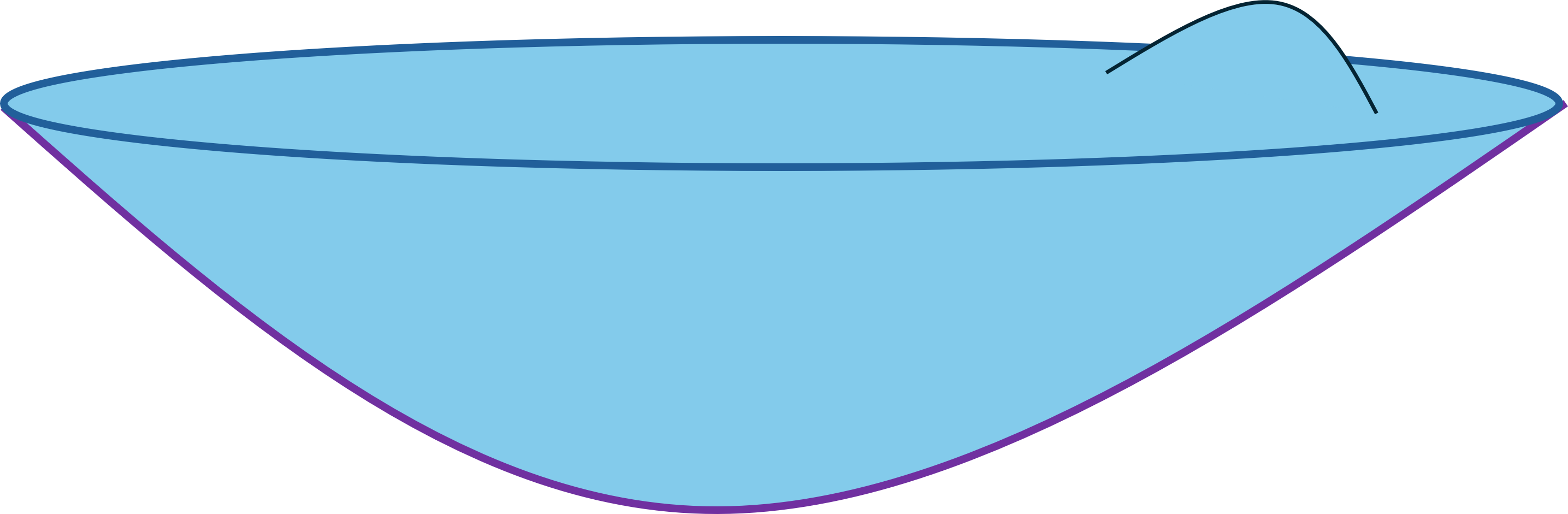}}
\newcommand{\monetwo}{\includegraphics[valign=m,scale=0.15]{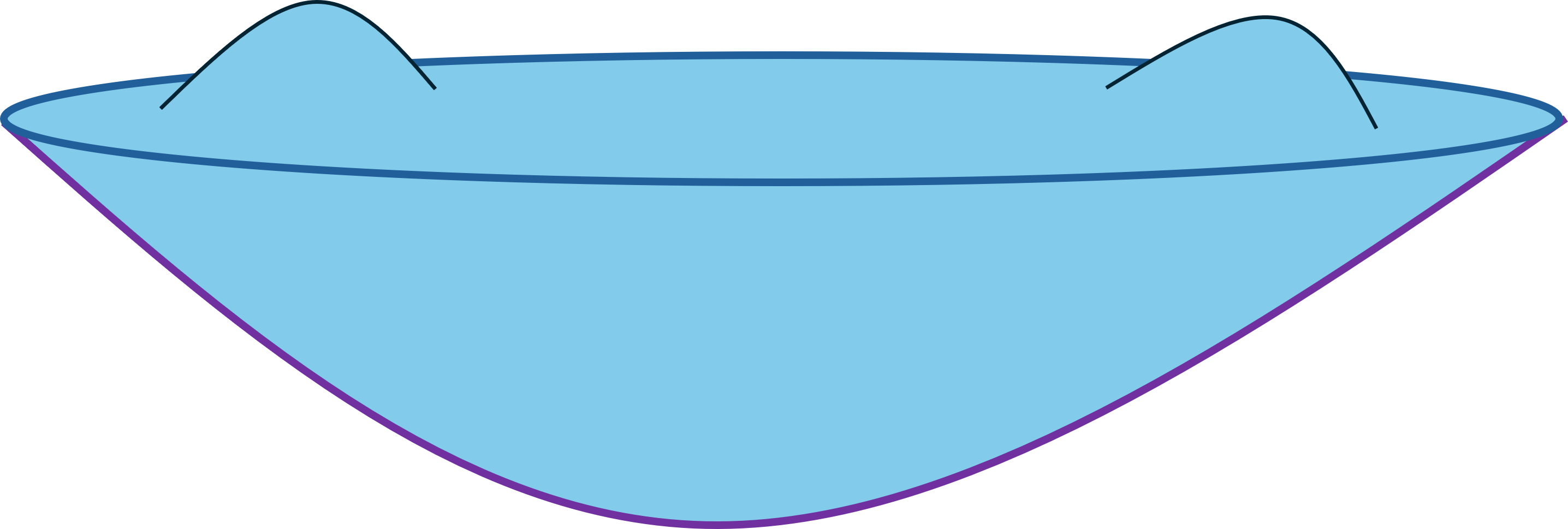}}

\usepackage[export]{adjustbox}
\usepackage{jheppub}
\usepackage{amssymb,amsmath,amsthm}
\usepackage[shortlabels]{enumitem}
\usepackage{braket}
\usepackage{breqn,hyperref}
\usepackage{comment} 

\usepackage{caption}
\usepackage{subcaption}
\usepackage{graphicx, xcolor, varwidth}
\usepackage{color}
\usepackage{float}

\theoremstyle{definition}

\theoremstyle{remark}

\DeclareMathOperator{\e}{\mathfrak{E}}

    \let\e=\varepsilon
   
\let\l=\lambda

 \let\G=\Gamma   \let\L=\Lambda

\newcommand{\Eqref}[1]{Eq.~\eqref{#1}}
\newcommand{\secref}[1]{Sec.~\ref{#1}}
\newcommand{\figref}[1]{Fig.~\ref{#1}}

\title{A Timelike Quantum Focusing Conjecture}
\author[]{Violet Concepcion and Kyle Ritchie}

\affiliation[]{Leinweber Institute for Theoretical Physics and Department of Physics,\\
University of California, Berkeley, California 94720, U.S.A.}

\emailAdd{violet\_concepcion@berkeley.edu}
\emailAdd{kyle\_ritchie@berkeley.edu}

\makeatletter
\gdef\@fpheader{\mbox{}}
\makeatother

\abstract{Recent proposals suggest that a notion of generalized complexity, analogous to generalized entropy, may be necessary for understanding the dynamics of holographic complexity in settings where quantum effects  are non-negligible, such as evaporating black holes.  Beginning with a notion of generalized complexity, we introduce a complexity-based quantum expansion for timelike geodesic congruences, and investigate the consequence of a timelike quantum focusing conjecture.  We find that for a suitable class of codimension-0 field theory complexity measures the timelike focusing condition implies a complexity-based quantum strong energy condition as well as a complexity bound which is analogous to the covariant entropy bound. }

\begin{document}
\maketitle
\section{Introduction}
Entanglement entropy and complexity have come to be recognized as two important measures of quantum information in gravitational systems.  The significance of entropy was first recognized in the context of black hole thermodynamics, where it was observed that in order for the second law of thermodynamics to be preserved, the area of causal horizons must bound the thermal entropy contained behind them \cite{Bekenstein:1973ur,BardeenCarterHawking1973,Hawking:1974sw}.  This led naturally to the concept of generalized entropy as the sum of the areas of all horizons plus the entropy of the matter on their exteriors, as well as to a generalized second law\cite{Bekenstein1974}.  More recently, the concept of generalized entropy has become best understood for arbitrary codimension-2 surfaces as the sum of their area (divided by $4G\hbar $) and the entanglement entropy of the semiclassical matter fields on a slice through their interior \cite{ Bousso:2015mna}.  In the context of holography, generalized entropy is central to understanding the flow of information in evaporating black hole systems and appears in the QES prescription as the AdS bulk dual to the entanglement entropy of CFT subregions \cite{Engelhardt:2014gca,Penington:2019npb,Almheiri:2019hni}.  It is now believed that these methods can be extended to include the von-Neumann entropy of bulk subregions \cite{Bousso:2022hlz,JensenSorceSperanza2023,Kaya:2025vof}.  Beyond the context of holography, focusing arguments have been used to understand the dynamics of generalized entropy for general codimension-2 surfaces.  In particular, simple assumptions about the quantum expansion of surfaces have been used to understand entropy bounds, such as the covariant entropy bound, and quantum energy conditions such as the qNEC, and used to predict the existence of quantum singularities \cite{Bousso:2015mna,Bousso_2023,Shahbazi-Moghaddam:2022hbw}. 

Nevertheless, it is well known that the entanglement dynamics are not sufficient to account for all of the known gravitational dynamics \cite{Susskind:2014moa}.  Computational complexity was first introduced in the context of gravity to understand the growth of the volume of the black hole interior after its entropy has saturated \cite{Susskind:2014rva, Stanford:2014jda}.  In that context, it was proposed that the holographic dual of CFT complexity is the volume of the maximal slice through the black hole spacetime.  Over the last decade, holographic complexity  has become better understood, culminating in many different refinements of the original complexity/volume proposal.  These include generalizations to subregions of the boundary CFT \cite{Alishahiha_2015,Agon:2018zso,Fan:2025moc}, generalizations to a broader class of candidate bulk duals beyond the volume \cite{Brown:2015bva,Couch:2016exn,Pedraza:2021fgp,Myers:2024vve}, and even an analysis of quantum corrections for different candidate complexity measures \cite{Emparan:2021hyr, AlBalushi:2020heq}.  Owing partially to the inherent ambiguities in quantifying the computational complexity for field theory state (e.g. choice of fundamental gate sets, cost functions, reference states), and partially to the relative recency of its introduction into quantum gravity, complexity lacks the robust theoretical understanding enjoyed by entanglement entropy.  For example, it is not known which (if any) of the broad class of existing candidate complexity duals is most relevant to holography. This is largely due to our lack of a unified framework for computing holographic complexity, such as the path integral techniques used for computing holographic entanglement entropy. 

It is widely understood that quantum corrections to the geometric part of any holographic complexity must be included to fully account for the dynamics beyond the classical limit.  The need for such corrections strongly suggests that complexity is dual not purely to geometric volume (or other proposed measures), but must include corrections from the complexity of the semiclassical bulk quantum fields. The authors of \cite{Concepcion:2026} argue that a notion of generalized complexity may be necessary to quantify the complexity dynamics of evaporating  black holes.  They demonstrate that a generalization of the CV proposal which includes bulk quantum corrections may be used to understand the appearance of the quantum extremal island, and they argue that the resulting complexity growth agrees with expectations from finite-dimensional qubit models.  This leads naturally to the question of whether such a notion of generalized complexity may be applicable to generic spacetime subregions. 

In what follows, we introduce a notion of generalized complexity for arbitrary partial slices $\Sigma$ as  the sum of their volume (in appropriate units) plus corrections from the semiclassical QFT complexity of the fields associated with $\Sigma$.  In order to study the dynamics of this generalized complexity, we introduce a notion of timelike quantum expansion, as the variation of the generalized complexity under local deformations of $\Sigma$ along an orthogonal timelike congruence through $\Sigma$.  We find that simple focusing assumptions lead to a complexity-based quantum strong energy condition, as well as a covariant bound on the complexity of fields by the geometric volume of their support.  We demonstrate that the focusing construction requires certain consistency conditions on the measure of QFT complexity.  For example, it must be codimension-0, and it must obey strong subadditivity.  We regard these conditions as a method for narrowing down large class of existing complexity measures, in the absence of reined mathematical control, to those natural for semiclassical gravity. 

In section 2, we introduce our definition of generalized complexity, and define a timelike quantum expansion.  In section 3, we investigate the consequences of the timelike quantum focusing assumption, and show that it entails a quantum energy condition, a complexity bound, and that its consistency requires strong-subadditivity of QFT complexity.  For clarity of presentation, in sections 2 and 3 we adapt the notion of the original QFC paper \cite{Bousso:2015mna} to the context of timelike congruences.  In section 4, we discuss a classical analogue of the derived entropy bound, an application to diagnosing quantum singularities, and then explore the relationship between the quantum Strong Energy Condition, its classical counterpart, and the cosmological constant.

\section{Definitions}
\subsection{Generalized Complexity}

\cite{Concepcion:2026} proposes that an appropriate extension of CV for incorporating islands is the following.  Given a boundary region $R$, the holographic complexity of this subregion is 
\begin{equation}\label{eq:C(R)}
    C(R) =\max_{\Sigma\in e(R)}\left(\frac{\rm{Vol}(\Sigma)}{\hbar G\ell}+ C_{QFT}(\Sigma)\right)
\end{equation}
where $\Sigma\in e(R)$ is a slice through the bulk component of the entanglement wedge of $R$, and $C_{QFT}(\Sigma)$ is the computational complexity of the semiclassical fields propagating in $e(R)$.  This motivates the introduction of a notion of generalized complexity for arbitrary subregions: 

Given a partial Cauchy slice $\Sigma$, we define its generalized complexity \begin{equation}\label{eq:C_gen}
    C_{gen}(\Sigma) = \frac{\rm{Vol}(\Sigma)}{\hbar G\ell} + C_{QFT}(\Sigma)
\end{equation} so that \Eqref{eq:C(R)} takes a form analogous to the QES prescription, but with a maximization over codimension-1 slices, rather than a minimization over codimension-2 surfaces:
\begin{equation}\label{eq:maxC_gen}
C(R) = \max_{\Sigma\in e(R)}C_{gen}(\Sigma)
\end{equation}

We do not yet make any specific claim about which complexity measure is used to quantify $C_{QFT}(\Sigma)$, nor the length scale $\ell$.  Many proposed complexity measures exist throughout the literature \cite{Dowling:2006tnk, Jefferson:2017sdb, Chapman:2017rqy,Balasubramanian:2022tpr}, but there is no reason any of them should be excluded at this step.  We take the point of view that there exists a measure (potentially many) for the complexity of the full quantum gravitational degrees of freedom associated with the slice $\Sigma$, and that this measure can be decomposed into a geometric piece and a piece associated with the semiclassical fields propagating on $\Sigma$.  For any given complexity measure, the geometric and QFT terms will be separately cutoff dependent, but we expect that their sum is cutoff independent.  Any rigorous justification will require a specific complexity measure.  The cutoff dependence of certain measures have been studied, for example, in \cite{Jefferson:2017sdb,Chapman:2017rqy,Bhattacharyya:2018bbv, Camargo:2019isp, Chen:2020nlj,Adhikari:2022whf}.

\subsection{Timelike Quantum Expansion}\label{sec:TQFC}

\begin{figure}[h]
    \centering
   \begin{subfigure}[b]{0.29\linewidth}
   \centering
\includegraphics[width=\linewidth]{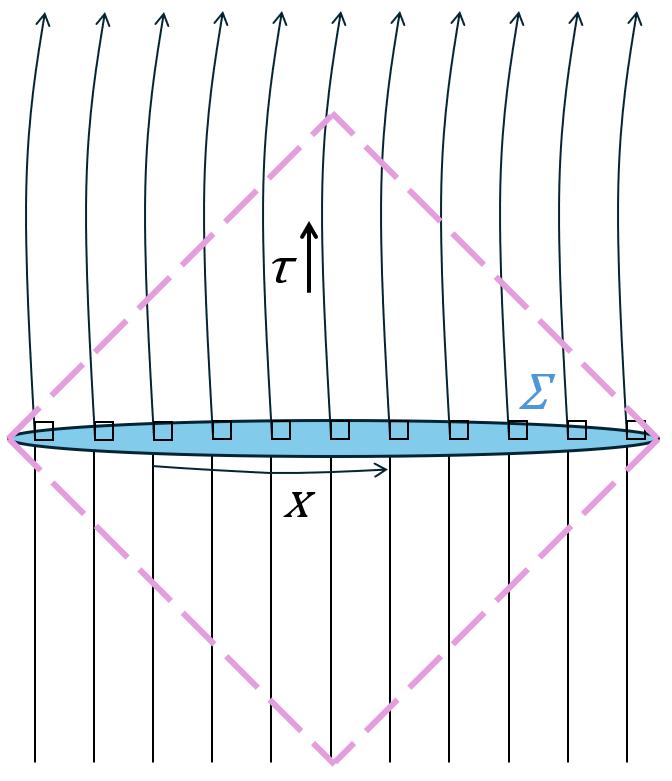}
\caption{Timelike Congruence}
\label{fig:Congruence}
    \end{subfigure}
    \hfill
    \begin{subfigure}[b]{0.35\linewidth}
        \centering
 \includegraphics[width=\linewidth]{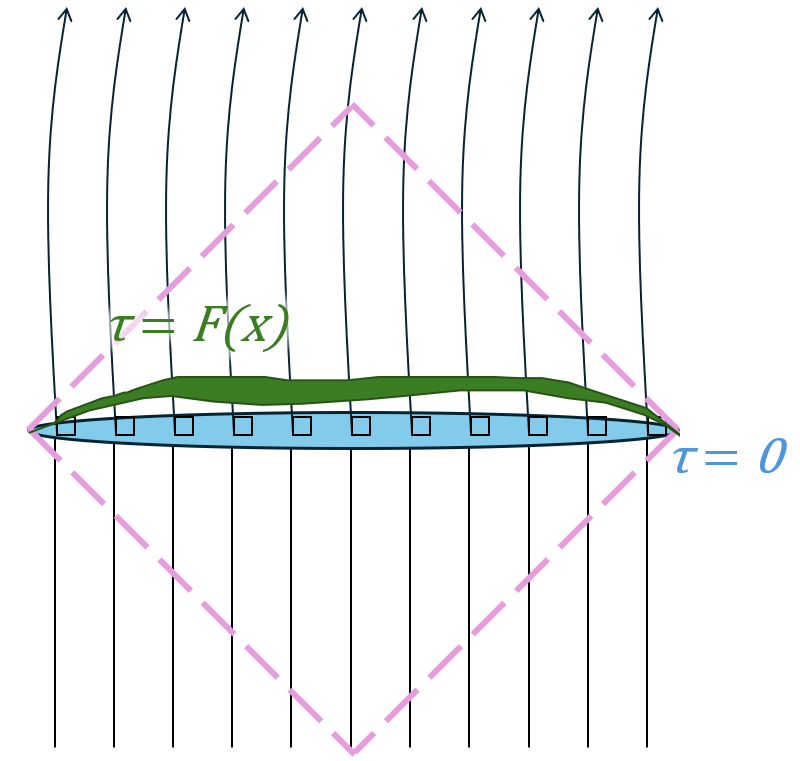}
    \caption{Finite deformation}
    \label{fig:finite_deformation}
    \end{subfigure}
    \hfill
     \begin{subfigure}[b]{0.34\linewidth}
        \centering
   \includegraphics[width=\linewidth]{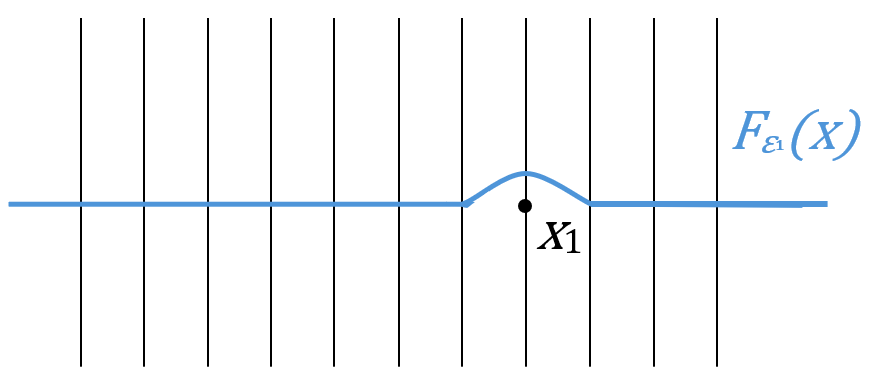}
    \caption{Pencil deformation}
    \label{fig:pencil_deformation}
    \end{subfigure}
    \caption{Panel (a) depicts a hypersurface orthogonal timelike congruence generated by a slice $\Sigma_0$.  In panel (b), deformations of the slice are defined by surfaces $\tau = F(x)$ where $\tau$ is the proper time along the timelike generator through $x\in\Sigma_0$.  Panel (c) depicts a local deformation along a ``pencil" of generators near a point $x_0$.}
    \label{fig:deformations}
\end{figure}

Consider a congruence $T$ of timelike geodesics which is torsionless, and which is hypersurface orthogonal to some partial Cauchy slice $\Sigma_0$.  We can introduce coordinates near the surface $\Sigma_0$ given by $(\tau, x)$ such that $\Sigma_0$ is parameterized by $(0,x)\in\Sigma_0$.  That is, $\tau$ is the coordinate orthogonal to the surface $\Sigma_0$, and labels the proper time along the geodesic that intersects $\Sigma_0$ at $(0,x)$.  See \figref{fig:deformations}. 

Other partial Cauchy slices through the domain of dependence of $\Sigma_0$ (i.e. those which share a boundary with $\Sigma_0$) can be obtained using a positive definite function $\tau = F(x)$, with the condition that $F(x)=0$ for $x\in\partial\Sigma_0$.  To ensure that the surface $\tau = F(x)$ is spacelike, we also restrict to the class of functions satisfying $|\nabla F(x)|^2<1$.  This condition ensures that the normal vector to $\Sigma$ remains timelike.  By defining partial Cauchy slices in this way, we can regard the generalized complexity as a functional of the embedding $F(x)$:
\begin{equation}
    C[F(x])] = \frac{\text{Vol}[F(x)]}{\hbar G\ell} + C_{QFT}[F(x)] 
\end{equation}

We may now consider deforming the slice in a neighborhood of a single generator through $x_1$.  That is, we consider a nearby slice of $T$ which differs from $\Sigma$ only in a neighborhood near $x_1$ having infinitesimal volume $\mathcal{V}$.  Such a deformation is described by a function 
$F_\epsilon(x) = F(x) + \epsilon \Delta_{x_1}(x)$
where $\Delta_{x_1}=1$ in a neighborhood of volume $\mathcal{V}$ around $x_1$ and is zero elsewhere\footnote{Note that $F_\epsilon(x)$ remains in our restricted class of functions, since $\epsilon$ may be chosen sufficiently small so that $F_\epsilon (x)=0=F(x)$ for $x\in\partial\Sigma_0$.  Additionally, to first order in $\epsilon$, $|\nabla F_\epsilon (x)|
^2<1$ so long as the condition holds for $F(x)$.}.  
We define the local deformation of $C_{gen}[F(x)]$ as follows: 
\begin{equation}
    \left.\frac{dC_{gen{}}}{d\epsilon}\right|_{x_1}\equiv\lim_{\epsilon\rightarrow 0}\frac{C_{gen}[F_{\epsilon}(x)]-C_{gen}[F(x)]}{\epsilon}
\end{equation}
We then define the timelike quantum expansion as 
\begin{equation}
    \Xi[F(x);x_1] \equiv \left.\lim_{\mathcal{V}\rightarrow 0}\frac{G\hbar\ell}{\mathcal{V}}\frac{d C_{gen}}{d\epsilon}\right|_{x_1}
\end{equation}
Equivalently, we may write the timelike quantum expansion as a functional derivative: 
\begin{equation}
    \Xi[F(x);x_1]\equiv \frac{G\hbar\ell}{\sqrt{g(x_1)}}\frac{\delta C_{gen}}{\delta F(x_1)}
\end{equation}
where $\sqrt{g}$ is the finite volume element of the metric restricted to $\Sigma$, and ensures that the functional derivative is taken per unit of coordinate-independent geometric volume. 

The factor of $\hbar G\ell$ is chosen so that in the limit that $\hbar\rightarrow 0$ the geometric term dominates the generalized complexity, and the quantum expansion reduces to the classical timelike expansion: 
\begin{equation}
    \lim_{\hbar\rightarrow 0}\Xi[F(x);x_1] = \lim_{\hbar\rightarrow 0}\frac{G\hbar\ell}{\sqrt{g(x_1)}}\frac{\delta C_{gen}}{\delta F(x_1)} = \frac{1}{\sqrt{g(x_1)}}\frac{\delta \text{Vol}[F(x)]}{\delta F(x_1)} = \vartheta[F(x);x_1]
\end{equation}

\section{Timelike Focusing and its Consequences}\label{sec:TQFC_Implications}

The timelike quantum focusing conjecture (TQFC) is the statement that the functional variation of the timelike quantum expansion is everywhere negative: 
\begin{equation}\label{eq:TQFC}
\frac{\delta}{\delta F(x_2)}\Xi[F(x);x_1]    \leq 0
\end{equation}
In what follows, we investigate three aspects of the TQFC inequality. 
\begin{enumerate}
    \item First, we consider the diagonal components of the TQFC, where $x_2=x_1$, and we demonstrate that it implies a quantum strong energy condition.
    \item Second, we consider the off-diagonal case, $x_2\neq x_1$, and we argue that for a suitable definition of $C_{QFT}$, the TQFC follows from strong subadditivity of $C_{QFT}$. 
    \item Third, we integrate the quantum expansion along non-expanding generators to obtain a complexity/volume bound which is analogous to the covariant entropy bound.  
\end{enumerate}

\subsection{A Quantum Strong Energy Condition}

The TQFC is a constraint on the second functional derivative of the generalized complexity:
\begin{equation}\label{eq:TQFC_2nd} \frac{\delta}{\delta F(x_2)}\frac{G\hbar\ell}{\sqrt{g(x_1)}}\frac{\delta C_{gen}}{\delta F(x_1)}\leq 0
\end{equation}
Consider first the diagonal case, where $x_2 = x_1$.  The timelike quantum expansion can be written, using a shorthand notation, as 
\begin{equation}
    \Xi = \vartheta + \frac{\hbar G\ell}{\mathcal{V}}\dot C_{QFT}
\end{equation}
where the `` $^\cdot$ " indicates a first variation along the (timelike) generator at $x_1$\footnote{We use a dot rather than a prime to emphasize that the variation plays the role of a time-derivative}.  The statement of the (diagonal) TQFC is that $\dot \Xi\leq 0$, which gives 
\begin{align}
    0\geq \dot\vartheta + \frac{G\hbar\ell}{\mathcal{V}}(\ddot C_{QFT} - \dot C_{QFT}\vartheta)
\end{align}

The classical timelike expansion obeys the Raychaudhuri equation \cite{Wald:1984rg}
\begin{equation}\label{eq:Raychaudhurri}
    \dot\vartheta = -\frac{1}{3}\vartheta^2 -\sigma^2+\omega^2 -  R_{\tau\tau}
\end{equation}
where $R_{\tau\tau} = R_{\mu\nu}u^\mu u^\nu$ and $u^\mu$ is the (unit) timelike vector field normal to $\tau =F(x)$.  $\sigma$ and $\omega$ denote the contracted shear and torsion tensors.  Then, using the trace-reversed form of the semiclassical Einstein equation\footnote{With this form of the Einstein equation, we have $R_{\mu\nu} = 8\pi (\langle T_{\mu\nu}\rangle -\frac{1}{2}\langle T\rangle g_{\mu\nu})$.  For a unit timelike vector $u^\mu$, we have $\langle T_{\mu\nu}\rangle u^\mu u^\nu - \frac{1}{2}\langle T\rangle g_{\mu\nu}u^\mu u^\nu = \langle T_{\tau\tau}\rangle  - \frac{1}{2}\langle T\rangle (-1) = \langle T_{\tau\tau}\rangle +\frac{1}{2}\langle T\rangle $.} and substituting into the TQFC inequality, we obtain 
\begin{equation}
    0\geq -\frac{1}{3}\vartheta^2 - \sigma^2-\omega^2 - 8\pi G\langle T_{\tau\tau}\rangle- 4\pi G \langle T\rangle + \frac{G\hbar\ell}{\mathcal{V}}(\ddot C_{QFT} - \dot C_{QFT}\vartheta)
\end{equation}
Note that we are free to choose a conrguence such that at the point $x_1$ the shear and torsion vanish, as well as the classical expansion: $\sigma^2=\omega^2=\vartheta^2=0$.  With this special choice of congruence, the $\hbar\rightarrow0$ limit yields the usual strong energy condition: 
\begin{equation}
    \langle T_{\tau\tau}\rangle +\frac{1}{2}\langle T\rangle \geq 0
\end{equation}
For nonzero $\hbar$, the same special choice of congruence gives an inequality which we refer to as the quantum strong energy condition (qSEC):
\begin{equation}\label{eq:qSEC}
    \langle T_{\tau\tau}\rangle + \frac{1}{2}\langle T\rangle \geq \frac{\hbar\ell}{8\pi\mathcal{V}}\ddot C_{QFT}
\end{equation}

This inequality is desirable for cases (such as AdS) in which the strong energy condition $T_{\tau\tau}+\frac{1}{2}T\geq 0$ is obeyed classically, but becomes locally violated once negative energy quantum fluctuations are allowed \cite{Epstein:1965zza,Fewster:2018pey}, yielding $\langle T_{\tau\tau}\rangle_{\Psi}+\frac{1}{2}\langle T\rangle _{\Psi}<0$.  The bound \Eqref{eq:qSEC}, if it holds, allows for negative energy fluctuations so long as they don't cause $\langle T_{\tau\tau}\rangle_{\Psi}+\frac{1}{2}\langle T\rangle _{\Psi}$ to become more negative than the second variation of the complexity of the state $\Psi$.  We also point out that, like the QNEC, the energy condition does not depend on the gravitational constant $G$, despite being derived from focusing conditions on geodesic congruences.  This suggests that the qSEC is a purely quantum inequality, and can be investigated without worrying about gravitational effects.  The inequality does however depend on the length parameter $\ell$, which we regard as a free parameter in our definition of quantum complexity.  In \secref{sec:disc} we discuss the plausibility of this inequality in general, and in cases where the strong energy condition is violated classically, such as de Sitter.

\subsection{Off-diagonal TQFC from SSA of Complexity}
In the previous section, the diagonal form of the TQFC was used to arrive at the qSEC.  In what follows, we show that the off-diagonal part, i.e. the case where $x_1\neq x_2$, follows from the strong sub-additivity (SSA) of complexity, once we restrict to a suitable class of $C_{QFT}$ measures.  In the context of entanglement entropy, SSA is the following property \cite{Lieb:1973cp}.  Suppose a state $\rho$ is supported on a partial Cauchy slice $\Sigma$ which can be decomposed into (possibly overlapping) subsystems $\Sigma_{A}$ and $\Sigma_B$ supporting states $\rho_A$ and $\rho_B$.  Then, 
\begin{equation}
    S(A\cup B) - S(A) - S(B) + S(A\cap B)\leq0
\end{equation}
Equivalently, for a tripartition of $\Sigma$ into (non-overlapping subregions)  $\Sigma_A\cup\Sigma_B\cup\Sigma_C$, 
\begin{equation}
    S(ABC) - S(AB) - S(BC) + S(C) \leq 0
\end{equation}

For measures of complexity that are slice-dependent, i.e., they depend only on the operators/fields acting on a slice $\Sigma$, it is unclear how the off-diagonal TQFC is related to SSA, if at all, since the functional variations vary the surface $\Sigma$ itself, rather than varying subregions of $\Sigma$.  However, we show below that if $C_{QFT}$ is taken to be in a class of codimension-0 measures, the off-diagonal QFC follows from SSA for codimension-0 subregions.  To demsonstrate this, we examine the form of the second $C_{gen}$ variation. 

Consider the case where $x_2\neq x_1$.  Since $\text{Vol}[F(x)]$ is the integral of a local functional of $F(x)$, and the factor $\sqrt{g}$ is evaluated at $x_1$, it follows that the off-diagonal second variation only receives a contribution from $C_{QFT}$.  Specifically, 
\begin{equation}
    \frac{\delta}{\delta F(x_2)}\frac{G\hbar\ell}{\sqrt{g(x_1)}}\frac{\delta C_{gen}}{\delta F(x_1)} = \frac{G\hbar\ell}{\sqrt{g(x_1)}}\frac{\delta^2 C_{QFT}}{\delta F(x_2)\delta F(x_1)}
\end{equation}
The second functional derivative can be written as follows: 
\begin{align}
    &\frac{\delta^2 C_{QFT}}{\delta F(x_2)\delta F(x_1)} =\\
    &\lim_{\mathcal{V}_1,\mathcal{V}_2\rightarrow 0}\frac{\sqrt{g(x_1)g(x_2)}}{\mathcal{V}_1\mathcal{V}_2}\lim_{\epsilon_1,\epsilon_2\rightarrow 0}\frac{C_{QFT}[F_{\epsilon_1,\epsilon_2}(x)] - C_{QFT}[F_{\epsilon_1}(x)] - C_{QFT}[F_{\epsilon_2}(x)]+C_{QFT}[F(x)]}{\epsilon_1\epsilon_2}
\end{align}
In the above expression, $\mathcal{V}_{1,2}$ are the volume elements at $x_1$ and $x_2$, and $\epsilon_{1,2}$ are the deformation parameters for the surface along the generators at $x_1$ and $x_2$.  The functions $F_\epsilon(x)$ are defined as before, e.g. 
\begin{equation}\label{eq:2ndvariation}
    F_{\epsilon_1,\epsilon_2}(x) = F(x) + \epsilon_1\Delta_{x_1}(x) + \epsilon_2\Delta_{x_2}(x)
\end{equation}
where the $\Delta_{x_i}$ are supported on $\mathcal{V}_i$, and are assumed to be non-overlapping.

 In the context of ordinary (null) quantum focusing, an identical expression to \Eqref{eq:2ndvariation} appears, but with $S_{out}$ replacing $C_{QFT}$, and with deformations taken along null generators rather than timelike ones.  A key insight was that the subregions of the (codimension-1) lightsheet above the slices deformed by the $\epsilon_i\Delta_{x_i}$s form a tripartition of the lightsheet.  This allows one to identify the quantities $S_{out}[F_{\epsilon_i}(y)]$ with the entanglement entropies of different lightsheet subregions.  In doing so, the non-positivity of the second variation becomes precisely the statement of strong subadditivity.  This suggests that $C_{QFT}[F(x)]$ ought to be defined in such a way that its variations allow us to form a tripartition of a (codimension-0) portion of the timelike congruence. 
 
 A natural way to proceed is to identify $C_{QFT}[F(x)]$ not as the complexity of constructing the state $\rho_{\Sigma}$ on $\Sigma$ using operators living on $\Sigma$, but as the complexity of preparing $\rho_\Sigma$ on the slice $\tau=F(x)$ from a reference state $\rho_{ref}$ on some earlier slice $\tau=F_{ref}(x)$. Indeed, complexity is fundamentally a count of operators and any operator which moves a state from one slice to another may be represented as a product of operators with support in the intermediate codimension-0 region. Optimizing over such preparation procedures allows us to associate $C_{QFT}[F(x)]$ with the codimension-0 region between $\Sigma_{ref}$ and $\Sigma$ rather than with $\Sigma$ alone\footnote{A simple example of such a procedure is the circuit associated with Hamiltonian evolution.  For a general reference state $\rho_{ref}$ additional operators may be necessary to prepare $\rho_{\Sigma}$.  In general, whether the Hamiltonian circuit is optimal depends on the precise details of the complexity measure, cost functions, etc.}.  Moreover, such a notion allows us to associate the terms in \Eqref{eq:2ndvariation} with subregions of the congruence beneath $\Sigma$ and its deformations, the union and intersection of which form a tripartion of a codimenison-0 region.
 
 \begin{figure}
    \centering
    \includegraphics[width=0.49\linewidth]{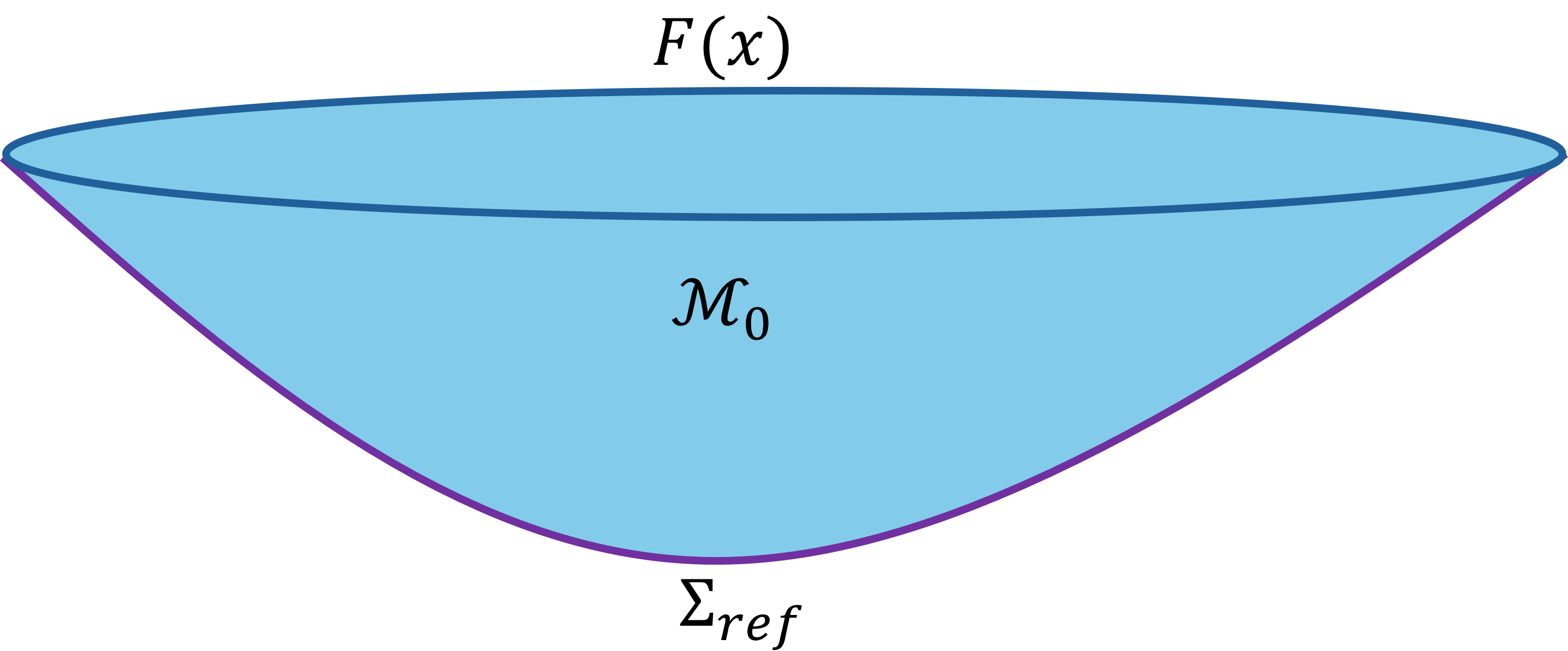}
    \includegraphics[width=0.49\linewidth]{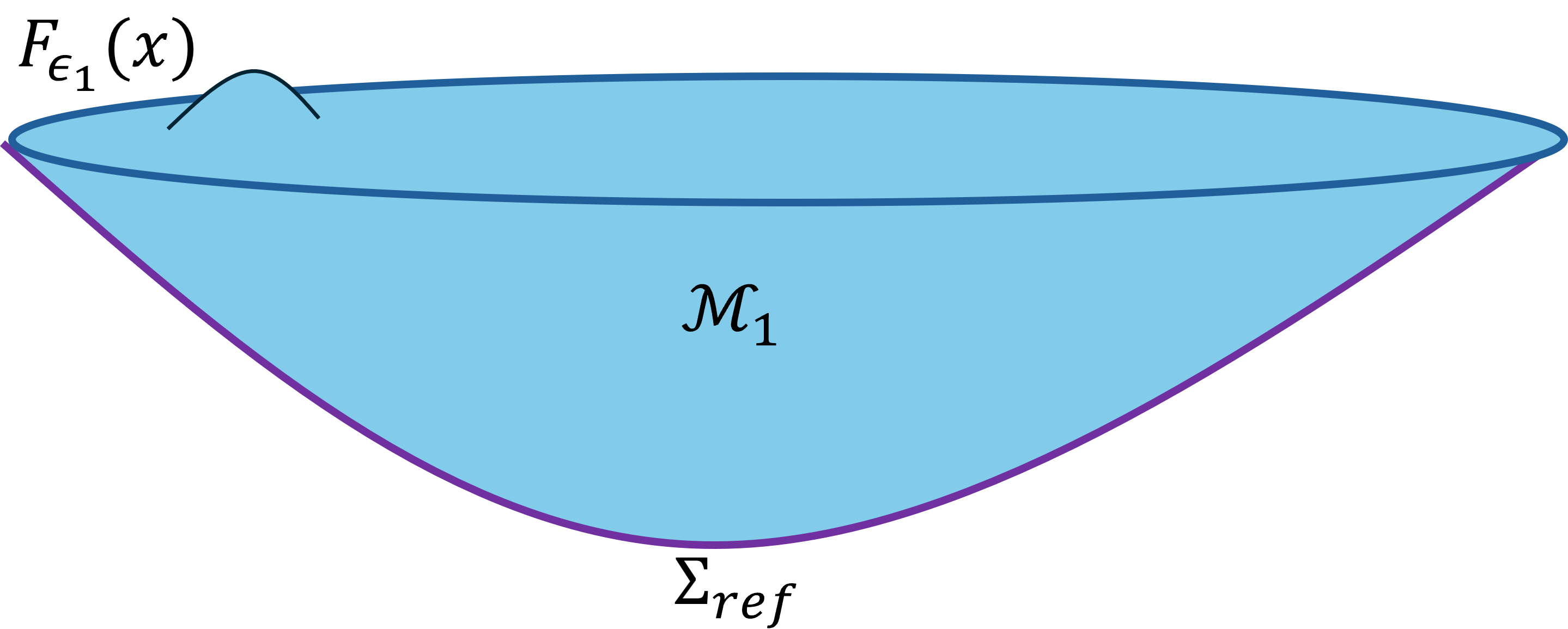}
    \includegraphics[width=0.49\linewidth]{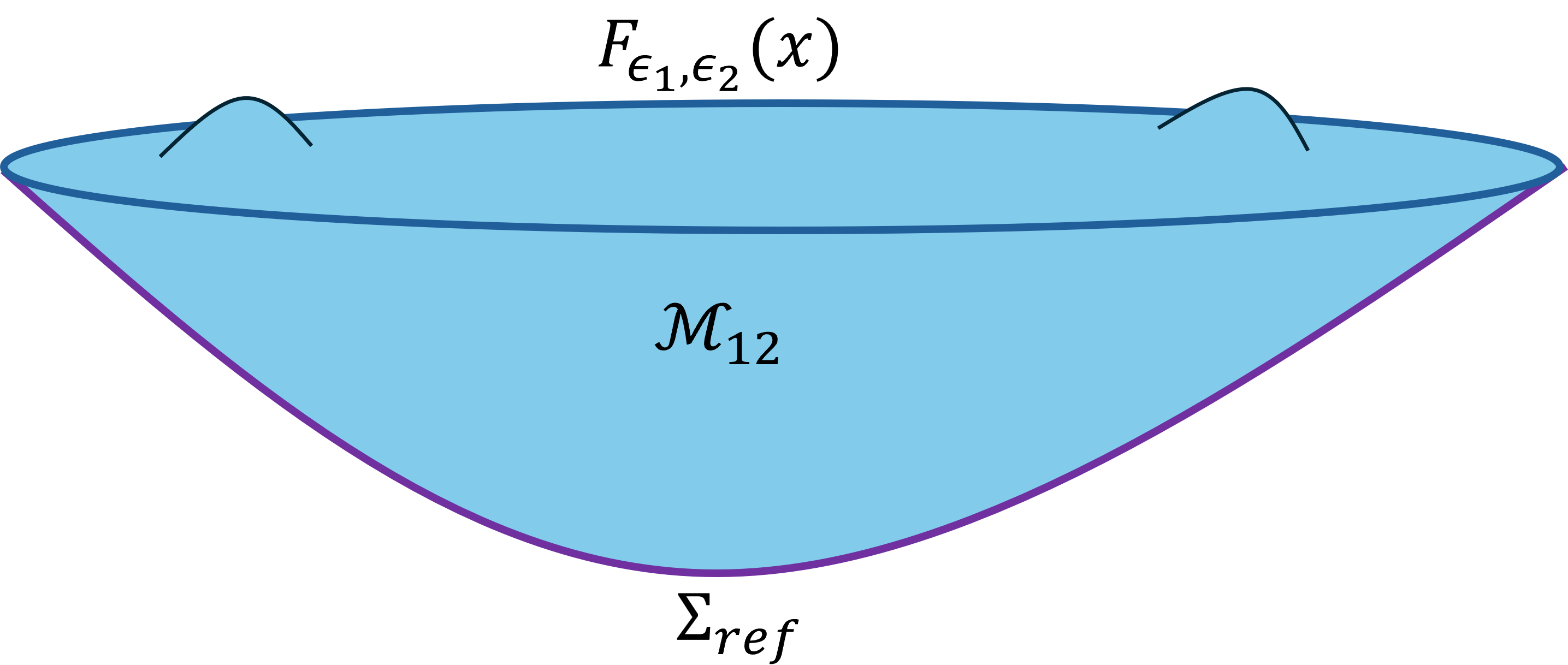}
    \includegraphics[width=0.49\linewidth]{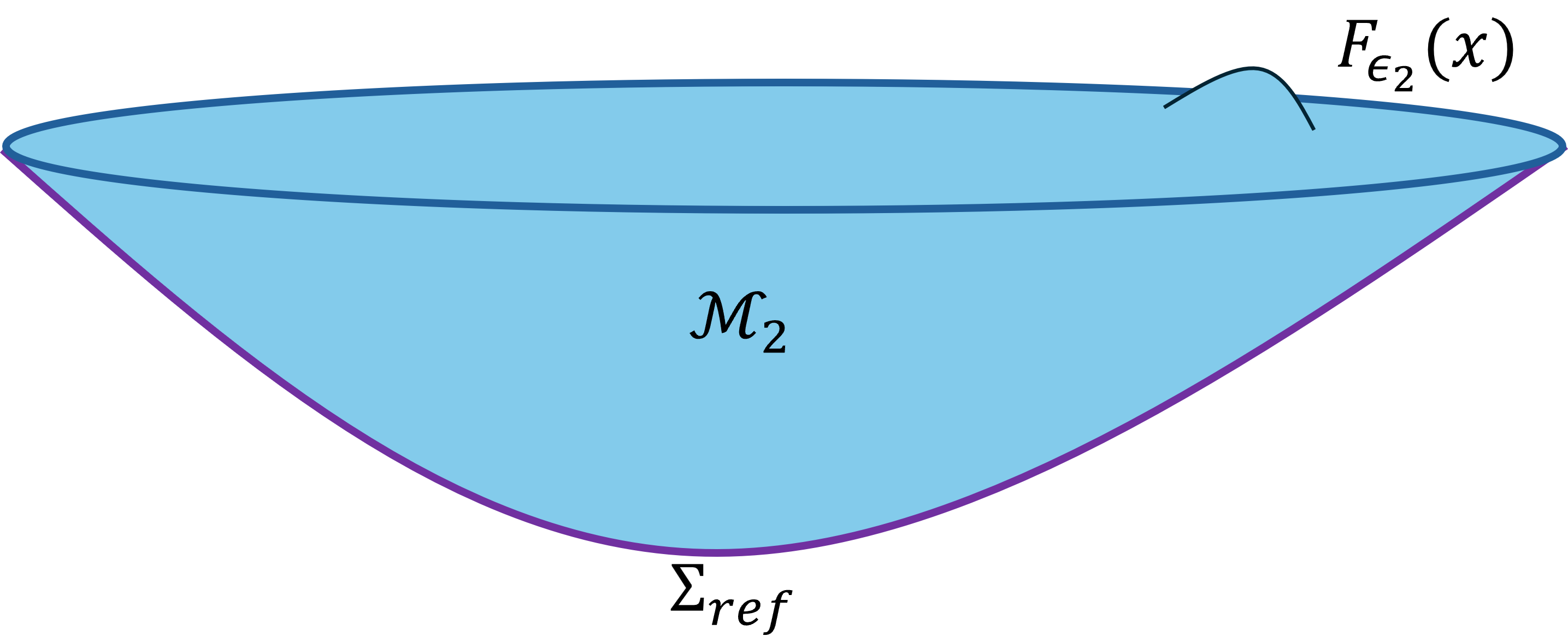}

    \caption{Depicted are the codimension-0 regions between the chosen reference slice $\Sigma_{ref}$ and the local deformations of the slice $\Sigma_0$.  The undeformed slice is defined by $\tau = F[x]$ (top left), and the deformations at $x_1$ and $x_2$ (right, bottom left) are implemented by $F_{\epsilon_1}(x)$ and $F_{\epsilon_2}(x)$.  The terms appearing in the second variation correspond to the complexities associated with these codimension-0 regions.} 
    \label{fig:SSA}
\end{figure}

Let us explicitly consider the terms in \Eqref{eq:2ndvariation}.  Denote the singly deformed slices $\tau = F_{\epsilon_i}(x)$ by $\Sigma_i$, the double deformed slice $\tau=F_{\epsilon_1,\epsilon_2}(x)$ by $\Sigma_{12}$, and the undeformed slice $\tau=F(x)$ by $\Sigma_0$.  Then, if we introduce some past reference state $\Sigma_{ref}$, we can denote the codimension-0 subregions of the congruence between $\Sigma_{ref}$ and $\Sigma_i$ by $M_{i} \equiv D^-(\Sigma_i)\cap D^+(\Sigma_{ref})$, $\mathcal{M}_{12}\equiv D^-(\Sigma_{12})\cap D^-(\Sigma_{ref})$, etc.  See \figref{fig:SSA}.  Explicitly, then, we have 
\begin{align}
    C_{QFT}[F(x)] &\equiv C(\mathcal{M}_0) = C\left(\mzero\right),\\
    C_{QFT}[F_{\epsilon_1}(x)]&\equiv C(\mathcal{M}_1)=C\left(\mone\right),\\
    C_{QFT}[F_{\epsilon_i}(x)]&\equiv C(\mathcal{M}_2)=C\left(\mtwo\right),\\ C_{QFT}[F_{\epsilon_1,\epsilon_2}(x)]&\equiv C(\mathcal{M}_{12})=C\left(\monetwo\right)
\end{align}
Finally, since $\mathcal{M}_{1}$, $\mathcal{M}_{2}$, and $\mathcal{M}_{12}$ are all obtained from $\mathcal{M}_0$ by future-directed defomations, it follows that $\mathcal{M}_0 = \mathcal{M}_1\cap\mathcal{M}_2$ and $M_{12}=\mathcal{M}_1\cup\mathcal{M}_2$, and the terms in \Eqref{eq:2ndvariation} become
\begin{align}
    C_{QFT}[F_{\epsilon_1,\epsilon_2}(x)] - C_{QFT}[F_{\epsilon_1}(x)] &- C_{QFT}[F_{\epsilon_2}(x)]+C_{QFT}[F(x)]\\
    =C(\mathcal{M}_1\cup\mathcal{M}_2)-C(\mathcal{M}_1)&-C(\mathcal{M}_2)+C(\mathcal{M}_1\cap\mathcal{M}_2)\label{eq:CSSA1}
\end{align}

Imposing nonpositivity on the last expression yields precisely the statement of strong subadditivity for the complexity associated with the overlapping regions $\mathcal{M}_1$ and $\mathcal{M}_2$.  The above expression can be written in a more familiar form if we tri-partition the subregion $\mathcal{M}_1\cup\mathcal{M}_2$ as follows: 
\begin{align}
    \mathcal
    A\equiv \mathcal{M}_1\cap(\mathcal{M}_1\cap\mathcal{M}_2) =&\includegraphics[valign=m,scale=0.2]{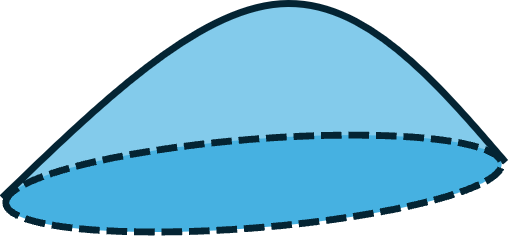}\hspace{1.5cm} \includegraphics[valign=m,scale=0.2]{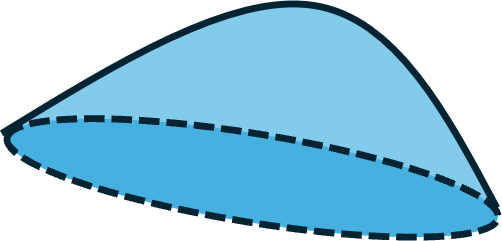}=\mathcal{B}\equiv\mathcal{M}_2\cap(\mathcal{M}_1\cap\mathcal{M}_2)\\
    \mathcal{C}\equiv\mathcal{M}_1\cap\mathcal{M}_2 =& \mzero
\end{align}
Then, the non-positivity of \Eqref{eq:CSSA1} can be stated as
\begin{equation}\label{eq:CSSA2}
    C(\mathcal{ABC})-C(\mathcal{AC}) - C(\mathcal{BC}) +C(\mathcal{C})\leq 0
\end{equation}
In either case, we find that the TQFC condition \Eqref{eq:2ndvariation} follows from the SSA inequality
\begin{align}
     C\left(\includegraphics[valign=m,scale=0.09]{figs/M12_new.png}\right) -C\left(\includegraphics[valign=m,scale=0.09]{figs/M1_new.png}\right) - C\left(\includegraphics[valign=m,scale=0.09]{figs/M2_new.png}\right)+ C\left(\includegraphics[valign=m,scale=0.09]{figs/M0_new.png}\right)\leq 0
\end{align}

This result reveals that the off-diagonal part of the focusing condition \Eqref{eq:TQFC} will hold in instances where the candidate codimension-0 complexity measure has the property of strong subadditivity.  There are many examples of codimenion-0 complexity measures throughout the literature, including the complexity=action proposal \cite{Brown:2015bva}, path-integral optimization \cite{Boruch:2020wax,Boruch:2021hqs}, and the infinite class of measures constructed in \cite{Belin:2022xmt}.  Different complexity measures obey various additivity inequalities, and it is not expected that every codimension-0 measure obeys SSA \cite{Caceres:2019pgf,Caceres:2018blh,Murugan:2026wss,Auzzi:2019vyh,Caceres:2025ypk,Nakajima:2026agi}.  We therefore regard the subadditivity condition as a useful guide for discerning which complexity measures are most relevant for focusing considerations. 

\subsection{A Covariant Complexity Bound}\label{sec:CCB}

We now investigate the consequence of integrating the timelike quantum expansion along non-expanding timelike generators, and arrive at a bound which we refer to as the covariant complexity bound (CCB).

Consider an initial slice $\Sigma$ defined by $\tau=0$, and some later slice $\Sigma'$ defined by $\tau=F(x)$ which is related to $\Sigma$ by deforming only along generators for which the timelike quantum expansion is negative.  In particular, we require that $F(x)=0$ for all points $x\in\Sigma$ where $\Xi[F(x);x]<0$.  Otherwise, $F(x)\geq 0$.  Then, since the nonpositivity of $\Xi$ along a generator $x$ implies that the generalized complexity is non-increasing along the generator, we see that the generalized complexity of $\Sigma'$ must be less than that of $\Sigma$.  That is, for this choice of $F(x)$, we have
\begin{equation}
    C_{gen}[F(x)]\leq C_{gen}[0],
\end{equation}
or equivalently, 
\begin{equation}
    C_{QFT}[\Sigma']+\frac{\text{Vol}(\Sigma')}{\hbar G\ell}\leq C_{QFT}[\Sigma]+\frac{\text{Vol}(\Sigma)}{\hbar G\ell},
\end{equation}
 and moreover 
 \begin{equation}
     C_{QFT}[\Sigma']-C_{QFT}[\Sigma] \leq \frac{\text{Vol}(\Sigma) - \text{Vol}(\Sigma')}{\hbar G\ell}
 \end{equation}
\begin{figure}[h!]
    \centering    \includegraphics[width=0.7\linewidth]{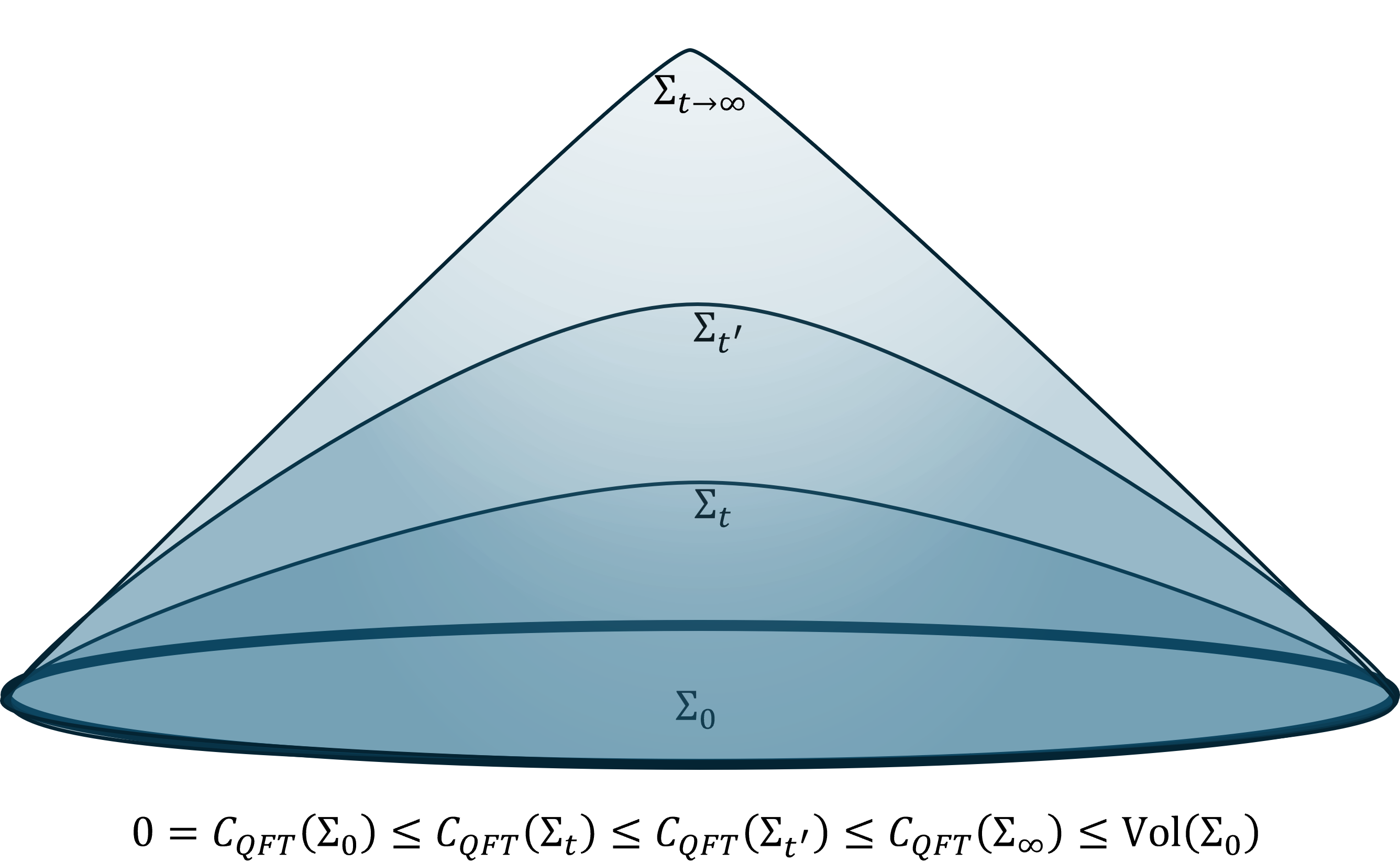}
    \caption{If the timelike quantum expansion is everywhere negative throughout the congruence, then the complexity associated with all future slices $\Sigma_t$ is bounded above by the volume of the initial slice $\Sigma_0$.}
    \label{fig:CCB}
\end{figure}
The above inequality is most illuminating in the following limit.  Suppose that we define $C_{QFT}$ as the complexity associated with Hamiltonian evolution through a codimension-0 region, and consider $\Sigma$ itself to be the reference slice supporting the initial field data.  Then, we have $C_{QFT}(\Sigma)=0$.  If $\Sigma'$ is chosen to be the future-most time-slice in any foliation of $D(\Sigma)$, then $\Sigma'$ approaches the future null boundary of $D(\Sigma)$, and its volume vanishes, i.e. $\text{Vol}(\Sigma')=0$.  In this limit, the inequality becomes
\begin{equation}
    C_{QFT}(\Sigma')\leq \frac{\text{Vol}(\Sigma)}{\hbar G\ell}
    \label{eq:CCB}
\end{equation}

Physically, this inequality states that if the fields on some compact slice $\Sigma$ are evolved as far as is causally allowed using the local Hamiltonian, then the complexity of the final state cannot exceed the volume of the slice $\Sigma$.  If we further make the reasonable assumption that $C_{QFT}$ is monotonically increasing along a foliation of slices $\Sigma_t$, with $\Sigma = \Sigma_0$, i.e. we assume that $t<t'\Rightarrow C_{QFT}(\Sigma_t)<C_{QFT}(\Sigma_{t'})$, then the inequality implies that 
\begin{equation}
    C_{QFT}(\Sigma_t)\leq \frac{\text{Vol}(\Sigma_0)}{\hbar G\ell}\quad \text{for all } t>0
\end{equation}
That is, the volume of a slice $\Sigma$ bounds the complexity of all future states obtainable from it by Hamiltonian evolution.  We discuss implications of this result in \secref{sec:disc}.

\section{Discussion}\label{sec:disc}
We have demonstrated that a notion of generalized complexity can be used to define a complexity-based timelike quantum expansion, and that focusing conditions on this quantum expansion imply a complexity-based quantum strong energy condition, as well as a new complexity bound.  We have also shown that the focusing condition (for non-overlapping deformations) follows from the strong subadditivity of complexity for codimension-0 complexity measures. 

We regard the focusing condition as a tool for narrowing the broad class of proposed complexity measures.  That is to say, we propose that the physically relevant complexity measure for semiclassical fields is one which obeys the neccesary focusing conditions, e.g. SSA.  We leave a detailed investigation of the focusing properties of particular complexity measure for future investigation. 

Below, we discuss the plausibility of the main consequences of focusing, as well as potential applications.  We provide evidence for the complexity bound \Eqref{eq:CCB} which doesn't rely on any notion of generalized complexity, and discuss how the complexity bound may be used to diagnose quantum singularities.  We then argue for the plausibility of the qSEC, and discuss how it may be preserved in spacetimes with positive cosmological constant where the SEC is violated. 

 \subsection{Classical Analogue of the Covariant Complexity Bound}

In \secref{sec:CCB}, we showed that the complexity of causally evolved fields from a slice $\Sigma$ is bounded by the volume of $\Sigma$, if the TQFC is assumed.  We now show that this bound is a covariant version of a complexity bound which can be deduced without any focusing considerations. 

Consider a region of space $\Sigma$ enclosed by a sphere of size $l$.  The amount of energy that can fit in this region without collapsing it into a black hole is bounded by \footnote{For example, the mass contained in a sphere of radius $r$ must be less than that of a Schwarzschild black hole of the same radius, i.e. $M\leq r/2G$. }
\begin{equation}
    E \lesssim \frac l G.
\end{equation}
Lloyd's bound \cite{Lloyd:2000cry} places a fundamental upper bound on the rate of complexity growth, 
\begin{equation}
    \dot{C} \lesssim \frac E \hbar
\end{equation}
where we have suppressed order one constants.  Given that the region $\Sigma$ has size $l$, the longest (future-directed) timelike geodesic contained in $\Sigma$'s domain of dependence has proper time of order $l$, i.e. $\Delta \tau \lesssim l$.  Then, for sufficiently small $l$ (or for sufficiently slowly varying $\dot C$), Lloyd's bound can be integrated to obtain 
\begin{equation}
    \Delta C \lesssim \frac{E}{\hbar}\cdot l\lesssim \frac {l^2}{G\hbar}
\end{equation}
Since in $d=3$ the volume of $\Sigma$ is $O(l^3)$, the above inequality can be written as
\begin{equation}
    \Delta C \lesssim \frac {\text{Vol}(\Sigma)}{\hbar G l}
\end{equation}
We regard the bound \Eqref{eq:CCB} as the covariant generalization of this statement, potentially indicating that the length parameter $\ell$ in the generalized complexity \Eqref{eq:C_gen} is state-dependent. 

\subsection{Diagnosing Singularities with Complexity Bounds}

The covariant complexity bound \eqref{eq:CCB} may be used to signal the existence of quantum singularities, in a manner analogous to \cite{Bousso_2023,Shahbazi-Moghaddam:2022hbw}.  To outline how this may be done, begin with a surface $\Sigma_0$ and consider future timelike deformations of $\Sigma_0$ which bring it to another surface $\Sigma$ in its domain of dependence.  Let's assume that the orthogonal congruence from $\Sigma_0$ used to perform deformations is future timelike trapped, or that everywhere in the congruence we have

\begin{equation}
    \vartheta< 0
\end{equation} 

Let us also assume deformations along the congruence only increase the quantum complexity, $C_{\text{QFT}}(\Sigma)$.  We will assume that by evolving along the slice we can bound the growth of complexity below by

\begin{equation}
    \frac{d}{d\tau} C_{\text{QFT}}(\Sigma) \geq \gamma
\end{equation}
for some positive $\gamma$.  Integrating this inequality gives

\begin{equation}
     C_{\text{QFT}}(\Sigma) \geq \gamma \tau + C_{\text{QFT}}(\Sigma_0)
\end{equation}
Comparing this to \Eqref{eq:CCB} we get

\begin{equation}
     \frac{\text{Vol}(\Sigma_0)}{\hbar G l} \geq \gamma \tau + C_{\text{QFT}}(\Sigma_0)
\end{equation}
or equivalently

\begin{equation}
     \tau \leq \frac1\gamma\left(\frac{\text{Vol}(\Sigma_0)}{\hbar G l}- C_{\text{QFT}}(\Sigma_0)\right)
\end{equation}

This inequality cannot hold beyond a finite proper time.  If the future development of $\Sigma_0$ were regular and timelike geodesically complete, one could continue the orthogonal flow to arbitrarily large $\tau$, contradicting the above inequality.  Physically, this says that finite regions cannot supply an infinite amount of complexity growth.  The singularity appears when the resources of a finite computational budget have run out. 
\subsection{The qSEC and the cosmological constant}

In spacetimes with a nonzero cosmological constant $\Lambda$, the strong energy condition is the statement that for any unit timelike vector field $u^\mu$ 

\begin{equation}
    \langle T_{\mu\nu}\rangle u^\mu u^\nu + \frac{1}{2}\langle T\rangle - \frac{\Lambda}{8\pi G}\geq0
\end{equation}

For $\Lambda <0$, the SEC holds in the vacuum (i.e. AdS), but can be violated if quantum fluctuations cause the energy expectation value to drop too low, i.e. if they cause $\langle T_{\mu\nu}\rangle u^\mu u^\nu + \frac{1}{2}\langle T\rangle \leq  \frac{\Lambda}{8\pi G}<0$.  However, the qSEC \Eqref{eq:qSEC}  becomes (upon including the cosmological constant),

\begin{equation}\label{eq:qSEC_CC}
     \langle T_{\tau\tau}\rangle + \frac{1}{2}\langle T\rangle  - \frac{\Lambda}{8\pi G}\geq \frac{\hbar\ell}{8\pi\mathcal{V}}\ddot C_{QFT}
\end{equation}
This inequality may still be preserved, provided that $\ddot C_{QFT}$ is sufficiently negative for states with SEC-violating energy fluctuations.  That is, if $\ddot C_{QFT}<0$, the qSEC allows for energy fluctuations to be more negative than the SEC allows for.  Moreover, since $C_{QFT}$ is explicitly state-dependent, it is plausible that $\ddot C_{QFT}$ provides a weaker (more negative) lower bound for states with greater negative energy fluctuations. 

For spacetimes with $\Lambda >0$, such as vacuum de Sitter, the SEC is violated even at the classical level, since $T_{\mu\nu}=0$ in the vacuum. This reduces the SEC inequality to $-\frac{\Lambda}{8\pi G}\geq 0$, which is explicitly violated in cases where $\Lambda >0$.  In order for the qSEC to hold while the SEC is violated, it must be the case that $\ddot C_{QFT} \lesssim -\Lambda$ even in the classical limit.  We speculate that this may be possible due to the cutoff dependence of $C_{QFT}$, and the natural IR cutoff inherent to observers $\Lambda >0$ spacetimes. 

Below, we sketch a framework for quantifying semiclassical complexity using a coarse-grained modular flow.  Within this framework, $\ddot C_{QFT}$ satisfies the negativity requirements outlined above, and scales with $\Lambda$ in a manner consistent with the requirements of the qSEC when $\Lambda >0$. 

\subsubsection*{Coarse-grained modular channels}

Consider a reference state on a slice through a domain of dependence $D$ in a generic spacetime.  An observer in a static patch will have a limit set on what observables they are actually able to measure.  We thus specify an algebra $A_\lambda$ of observables in the diamond below a cutoff $\l$ which keeps only observables with modular frequencies $|\omega|\leq \l$.  We also introduce a tolerance $\e >0$ within which nearby states are identified in an appropriate norm.  The resulting Hilbert space is finite-dimensional, and we suppose that there is a finite set of gate operators $G\subset A_\l$ which can be used to construct arbitrary semiclassical states with accuracy $\e$. 

This allows us to define a coarse grained modular channel as follows.  Let $U_D = \exp\{-i Ks\}$ denote the modular operator, where $K$ is the modular hamiltonian and $s$ is the modular parameter.  An operator $X\in A_\l$ evolves under modular evolution via $X_s = U_D(-s) XU_D(s)$.  We define its evolution in the coarse-grained modular channel as
\begin{equation}
X^{(\l)}_s = P_{A_\l}(U_D(-s)XU_D(s)),\quad   X\in A_\l
\end{equation}
where $P_{A_\l}$ is an operator that projects only on observables below the cutoff $\l$.  This is necessary since exact modular evolution may evolve $X_s$ outside of $A_\l$.  We can then define an effective channel complexity as the minimal number of gates needed to approximately construct $X^{(\l)}$ from $X$ within the $\e$ tolerance.  Formally, if $W=g_1\cdots g_N$ is a channel constructed of gates $g_i\in G$ such that $|X^{(\l)}_s - W^\dagger X W| \leq \e$, then the effective channel complexity is the minimal $N$ over all such channels $W$: 
\begin{equation}
    C_c(s,\e,\l) = \min_{W} N(W)
\end{equation}

Once the effective Hilbert space is finite dimensional and one works at fixed nonzero tolerance $\e$, the set of distinguishable channels is compact up to an $\e$-net, so the approximate complexity is bounded above by some finite $C_{max}$ depending on the regulator, gate set, and tolerance
\cite{Dawson:2005blj,Oszmaniec:2022srs}.  This means that 
\begin{equation}
0\leq C_c(s) \leq C_{max}< \infty
\end{equation}

Following other examples in the literature \cite{BrownSusskind2017SecondLaw,Zhao2018Uncomplexity}, we introduce an ``uncomplexity reservoir" as the difference between the complexity and the maximum complexity
\begin{equation}
u(s) \equiv C_{max}-C_c(s)
\end{equation}
which is always positive.  Since we have coarse grained, we do not track microscopic recurrences and only ask if the next increment of modular flow is taking the retained observables into a new channel class or into one already explored.  When $u(s)$ is large, nearly every step explores a new channel.  Late in the evolution, the growth slows down and the system saturates.  We can model the evolution of the coarse-grained complexity with a simple saturation ansatz
\begin{equation}
\frac{d}{ds} C_c(s) = \G u(s) = \G (C_{max} - C_c(s))
\end{equation}
where $\G$ is some coefficient that describes the effective mixing rate.  Generically, $\Gamma$ will may depend on the length scale of the chosen geometry, as well as the cutoff scale $\lambda$.  Solving this differential equation gives
\begin{equation}
\dot{C}_c(s)= \G (C_{max} - C_c(0))e^{-\G s}
\end{equation}
and therefore
\begin{equation}
\ddot{C}_c(s)= -\G^2 (C_{max} - C_c(0))e^{-\G s}<0
\end{equation}
 
\subsubsection*{The classical limit in AdS and dS}
If one assumes that the timelike congruence can be chosen so that $\frac{d^2}{d\tau^2}$ and $\frac{d^2}{d s^2}$ only differ by a positive factor, the above inequality gives the negative concavity required by the qSEC. In the case of AdS, if the mixing rate $\Gamma$ goes like $\lambda$ as we would expect, i.e. $\Gamma\sim\lambda$, then in the limit  $\lambda\rightarrow 0$ where all QFT modes are coarse-grained over, we have $\Gamma\rightarrow0$ and hence $\ddot C \rightarrow0$, thus reducing the qSEC to the SEC. 

In the case of the dS static patch vacuum, it remains true that $\ddot C\leq 0$ under the earlier assumptions. However, flowing the observer's cutoff scale into the IR corresponds to $\lambda\to\ell_{dS}^{-1}$, rather than $\lambda\rightarrow 0$, since the static patch radius is the largest possible length scale in de Sitter. In this limit, we thus have that $\Gamma\sim1/\ell_{dS}$. To estimate the saturation complexity, we take the effective number of fundamental qubits $N_{\text{eff}}$ needed to describe the static patch to scale with the dS entropy.
\begin{equation}
N_{\text{eff}} \sim S_{dS} = \frac{A_{dS}}{\hbar G}=4\pi\frac{\ell_{dS}^2}{\hbar G}
\end{equation}
The static patch volume is
\begin{equation}
V_{patch} \sim \frac{4}{3}\pi\ell_{dS}^3,
\end{equation}
so the corresponding maximum coarse-grained complexity density in the IR scales like
\begin{equation}
c_{max} = \frac{C_{max}}{V_{patch}} \sim \frac{N_{\text{eff}}}{V_{patch}}\sim\frac3{\hbar G\ell_{dS}}
\end{equation}
Plugging this into our saturation model from before gives
\begin{equation}
\ddot c\sim-\G^2c_{max}=-\frac{3}{\hbar G\ell_{dS}^3}
\end{equation}
Finally,
\begin{equation}
\frac{\hbar\ell_{dS}}{8\pi V} \ddot C\sim-\frac{3\hbar \ell_{dS}}{8\pi\hbar G\ell_{dS}^3}=-\frac{3}{8\pi G\ell_{dS}^2}= - \frac{3\Lambda}{8\pi G}=-\rho_\L
\end{equation}
where $\rho_\L$ is precisely the vacuum energy density in $dS$.  Comparing this with the discussion below \Eqref{eq:qSEC_CC} reveals that the semiclassical complexity is at precisely the scale needed for the qSEC to hold in vacuum de Sitter.  This suggests that $\ddot C$ may remain negative in the classical limit and may exist as a classical observable. We leave a detailed analysis of the relationship between effective QFT complexity and the cosmological constant to future work.  

These conclusion are possible because $C_{gen}(\Sigma) = \frac{\text{Vol}(\Sigma)}{\hbar G \ell } + C_{QFT}(\Sigma)$ is intrinsically a semiclassical object.  In particular, for the qSEC to make sense in de Sitter-like spacetimes, the renormalized theory must contain finite infrared complexity dynamics.  Since the exact vacuum state in a causal diamond is stationary under exact modular flow, this finite contribution cannot naturally be interpreted as the second derivative of exact state complexity.  Exact modular flow leaves the reduced vacuum state invariant, whereas coarse-grained modular flow can display relaxation and saturation, making a negative second derivative of an effective complexity observable possible.  Therefore, it should be associated with this renormalized and coarse-grained notion of modular complexity, or an equivalent effective complexity observable attached to the observer algebra of the static patch.  If such an observable is nonzero, then the static patch exhibits infrared complexity dynamics even in the vacuum.  This suggests a maximum finite coarse-grained complexity capacity for the patch, which is naturally compatible with the presence of a horizon, a temperature, and a finite de Sitter entropy.  A similar interpretation should hold in black-hole spacetimes, where the relevant observer algebra is again associated with a thermal horizon subsystem rather than a globally pure vacuum state. 

\acknowledgments

We thank Mathew Self, Liz Wildenhain, Hugo Marrochio, and Yasunori Nomura for helpful discussions on early versions of this work. This work was supported by the U.S. Department of Energy, Office of Science, Office of High Energy Physics under QuantISED award DE-SC0019380 and contract No.\ DE-AC02-05CH11231.

\bibliographystyle{JHEP}
\bibliography{refs.bib}
\end{document}